\documentclass[journal,times,onecolumn,12pt]{IEEEtran}
\usepackage{cite}

\usepackage[margin=1in]{geometry}
\usepackage{graphicx} 
\usepackage{caption}
\usepackage{subcaption}
\usepackage{float}
\usepackage{subfloat}
\graphicspath{{./Figures/}}
\usepackage{epsfig}
\usepackage{epstopdf}

\usepackage[american]{circuitikz}
\usepackage{tikz}
\usetikzlibrary{arrows}

\usepackage{siunitx}
\usepackage{booktabs} 
\usepackage{multirow}

\usepackage[cmex10]{amsmath}

\usepackage{array}

\begin{document}

\title{Analysis on One-Stage SSHC Rectifier for Piezoelectric Vibration Energy Harvesting}

\author{Sijun~Du
\thanks{The author is with the Department of Electrical Engineering and Computer Sciences, University of California at Berkeley, Berkeley, CA, 94720, USA.} 
}

\maketitle


\begin{abstract}
Conventional SSHI (synchronized switch harvesting on inductor) has been believed to be one of the most efficient interface circuits for piezoelectric vibration energy harvesting systems. It employs an inductor and the resulting RLC loop to synchronously invert the charge across the piezoelectric material to avoid charge and energy loss due to charging its internal capacitor ($C_P$). The performance of the SSHI circuit greatly depends on the inductor and a large inductor is often needed; hence significantly increases the volume of the system. An efficient interface circuit using a synchronous charge inversion technique, named as SSHC, was proposed recently. The SSHC rectifier utilizes capacitors, instead of inductors, to flip the voltage across the harvester. For a one-stage SSHC rectifier, one single intermediate capacitor ($C_T$) is employed to temporarily store charge flowed from $C_P$ and inversely charge $C_P$ to perform the charge inversion. In previous studies, the voltage flip efficiency achieves 1/3 when $C_T = C_P$. This paper presents that the voltage flip efficiency can be further increased to approach 1/2 if $C_T$ is chosen to be much larger than $C_P$. 

\end{abstract}

\section{Introduction} \label{1907_intro}

Along with the development of Internet of Everything, wireless sensing networks (WSN) act as essential roles interconnecting between the real world and the Internet. Although electrochemical batteries have remained the primary energy sources for such systems due to the high energy density, certain sensors and sensor systems require to operate over significant periods of time much longer than the lifetime of electrochemical batteries. Battery usage may be both impractical and costly due to the requirement for periodic re-charging and/or replacement \cite{Vullers2009,  Rezaei2016atdom, Belleville2010, Niu2015auscsd}. In order to address this challenge and extend the operational lifetime of wireless sensors, there has been an emerging research interest on harvesting ambient vibration energy. 

Piezoelectric materials are widely used in vibration energy harvesters (VEH) as mechanical-to-electrical transducers due to their relatively high power density, scalability and compatibility with conventional integrated circuit technologies \cite{Liangjunrui2012}. MEMS technology is also widely used for piezoelectric vibration energy harvesters and different kinds of resonant sensors \cite{luo2016daaombf}. A typical piezoelectric VEH can provide a power density of around 10 - 500 \si{\micro\watt\per\square\centi\meter}, which sets a significant constraint on designing the associated power conditioning interface circuit. The interface circuit not only must consume ultra low power, but it also should be able to recover the power as effectively as possible from the piezoelectric transducer (PT). Full-bridge rectifiers are widely used in commercial energy harvesting systems due to their simplicity and stability; however, they set high threshold voltages for the generated energy to be extracted by the circuit \cite{Szarka2012, Liu2012, Blystad2010pmehs}. 

\begin{figure}
\centering
\begin{subfigure}{0.49\textwidth}
  \centering
\begin{tikzpicture}[american currents, scale=0.71, transform shape]
\draw (2.5,1) node[anchor=north]{$V_N$} -- (3-0.5,3)  (3-0.5,5)--(3-0.5,7) node[anchor=south]{$V_P$};
\draw (1.5,3) -- (3.5,3) (1.5,5) -- (3.5,5);
\draw (1.5,3) to[I] (1.5,5) (2.5,3)node[circ]{} to[C] (2.5,5)node[circ]{}  (3.5,3) to[R] (3.5,5);
\node at (1.1,4.6) {$I_P$};
\node at (2.1,4.6) {$C_P$};
\node at (3.1,4.6) {$R_P$};
\draw (3-0.5,7) -- (6,7) to [D] (8,7);
\draw (8,1) to [D] (6,1) -- (3-0.5,1);
\draw (5.5,1) to [short,*-] (5.5,5) -- (6,5) to [D, l=$V_D$] (8,5) to [short,-*] (8,7);
\draw (8,1) to[short,*-] (8,3) to [D] (6,3) to [short,-*] (6,7);
\draw (9.5,1) to [C=$C_S$] (9.5,7);
\draw (8,7) -- (9.5,7) (8,1) -- (9.5,1);
\draw (8,1) node[ground]{};
\draw [->] (4.7,1.1) -- (4.7,6.9) ;
\draw (4.7,6.5) node[anchor=west]{$V_{PT}$} (9.5,5)node[anchor=east]{$V_S$};
\end{tikzpicture}
\caption{The equivalent circuit of a piezoelectric VEH connected to a full-bridge rectifier}
\label{fig:1907_ecoapvecfr}
\end{subfigure}
\begin{subfigure}{0.49\textwidth}
  \centering
\begin{tikzpicture}[american currents, scale = 0.7, transform shape]
\draw [thick, ->] (3,0) -- (3,8) node[anchor=east]{$I_p$};
\draw [thick, ->] (3,0) -- (3,4) node[anchor=east]{$V_{PT}$};
\draw [thin,->] (3,6)--(11.5,6) node[anchor=west]{$time$};
\draw [thin,->] (3,2)--(11.5,2) node[anchor=west]{$time$};

\draw[thick] (3,6) sin (4,7.5) cos (5,6) sin (6,4.5) cos (7,6)
          sin (8,7.5) cos (9,6) sin (10,4.5) cos (11,6);
          
\draw [dotted] (3,3.5) -- (10,3.5) node[anchor=south] {$V_S + 2V_D$} ;
\draw [dotted] (3,0.5)  -- (10,0.5) node[anchor=north] {$-(V_S + 2V_D)$};

\draw [thick](3,0.5) parabola (4.3,3.5);
\draw [thick](4.3,3.5) -- (5,3.5);
\draw [thick](5,3.5) parabola (6.3,0.5);
\draw [thick](6.3,0.5) -- (7,0.5);
\draw [thick](7,0.5) parabola (8.3,3.5);
\draw [thick](8.3,3.5) -- (9,3.5);
\draw [thick](9,3.5) parabola (10.3,0.5);
\draw [thick](10.3,0.5) -- (11,0.5);

\draw [dotted] (4.3,8) -- (4.3,0);
\draw [dotted] (5,8) -- (5,0);
\draw [dotted] (6.3,8) -- (6.3,0);
\draw [dotted] (7,8) -- (7,0);
\draw [dotted] (8.3,8) -- (8.3,0);
\draw [dotted] (9,8) -- (9,0);
\draw [dotted] (10.3,8) -- (10.3,0);

\end{tikzpicture}
\caption{Representative waveforms for the current ($I_P$) and voltage ($V_{PT}$)}
\label{fig:1907_rwftcipavp}
\end{subfigure}
\caption{Full-bridge rectifier for piezoelectric VEH and the associated waveforms }
\label{fig:1907_fbrfpvataw}
\end{figure}
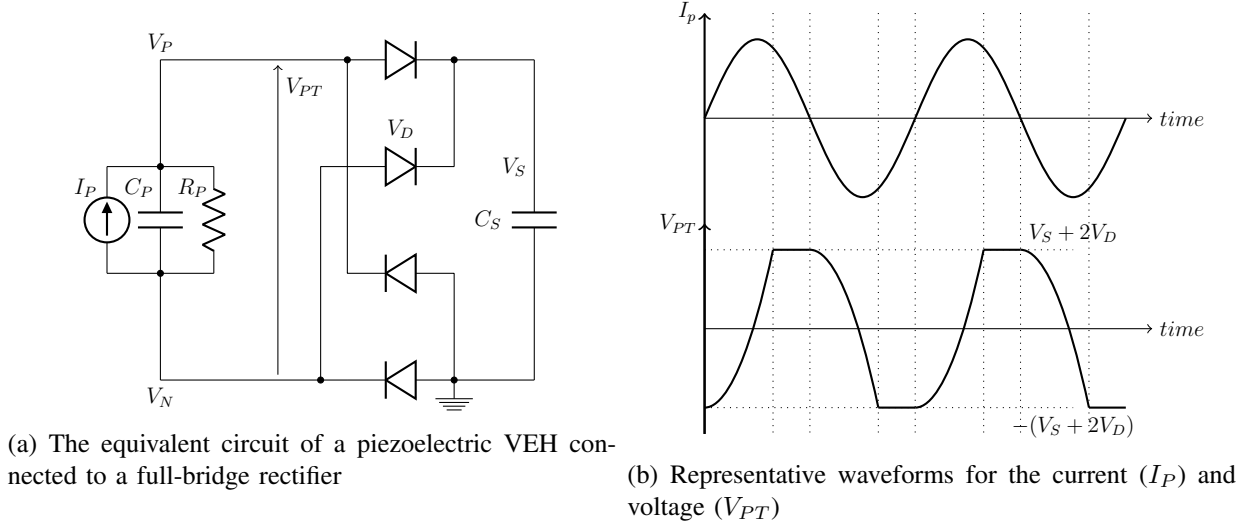

Fig. \ref{fig:1907_fbrfpvataw} shows a full-bridge rectifier connected with a piezoelectric transducer (PT) and the associated waveforms. While vibrating at or close to its resonance, a piezoelectric VEH can be modeled as a current source $I_P$ connected in parallel with a plate capacitor $C_P$ and a resistor $R_P$. In order to transfer the generated energy from the PT to the storage capacitor $C_S$, the voltage across the PT $V_{PT}$ should attain $V_S + 2 V_D$ or $-(V_S + 2V_D)$. Hence the energy used to charge the internal capacitor $C_P$ from $V_S + 2 V_D$ to $-(V_S + 2V_D)$ (or vice-versa) is wasted. The wasted charge in a half vibration cycle can be expressed as:

\begin{equation} \label{eq:1907_fbrtfwepz}
\centering
Q_{wasted} = 2 C_P ( V_S + 2V_D)
\end{equation}

Therefore, the peak-to-peak open-circuit voltage generated by the PT, $V_{pp(open)}$, should be greater than $2(V_S + 2V_D)$ to make sure that there is remaining charge flowing into $C_S$ after $Q_{wasted}$ is wasted. In large excitation levels, the charge loss $Q_{wasted}$ may take a small part of the total charge generated by the PT. However, in small excitation levels where $V_{pp(open)}$ is below or marginally higher than $2(V_S + 2V_D)$, there will be no charge or very little charge can be transferred to $C_S$. The waveforms in Fig. \ref{fig:1907_fbrfpvataw} illustrate the charge loss due to charging $C_P$ from $\pm(V_S + 2 V_D)$ to $\mp(V_S + 2V_D)$.

\section{SSHC interface circuit} \label{1907_sshiic}

In order to improve the power efficiency and minimize the charge waste due to charging $C_P$, many active interface circuits have been reported, including MPPT (maximum power point tracking) \cite{Szarka2013upfabrfe, Dongwon2014jasi, Sanchez2016apsrfp}, active rectifications, etc \cite{Ramadass2010, Lefeuvre2017aopeh, Aktakka2014}. Among all interface circuits for piezoelectric VEH, SSHI (Synchronized Switch Harvesting on Inductor) rectifier, proposed by Badel et al \cite{badel2005efficiency}, is one of the most efficient circuits with nearly no charge waste (if the inductor is chosen large enough). It performs synchronous charge inversion on $C_P$ through an RLC system using an inductor controlled by synchronized switches. However, the limitation of the SSHI rectifier is the need for a large off-chip inductor. This section presents the design of a rectifier, named as synchronized switch harvesting on capacitor (SSHC), which performs the energy extraction using switched capacitors instead of inductors.

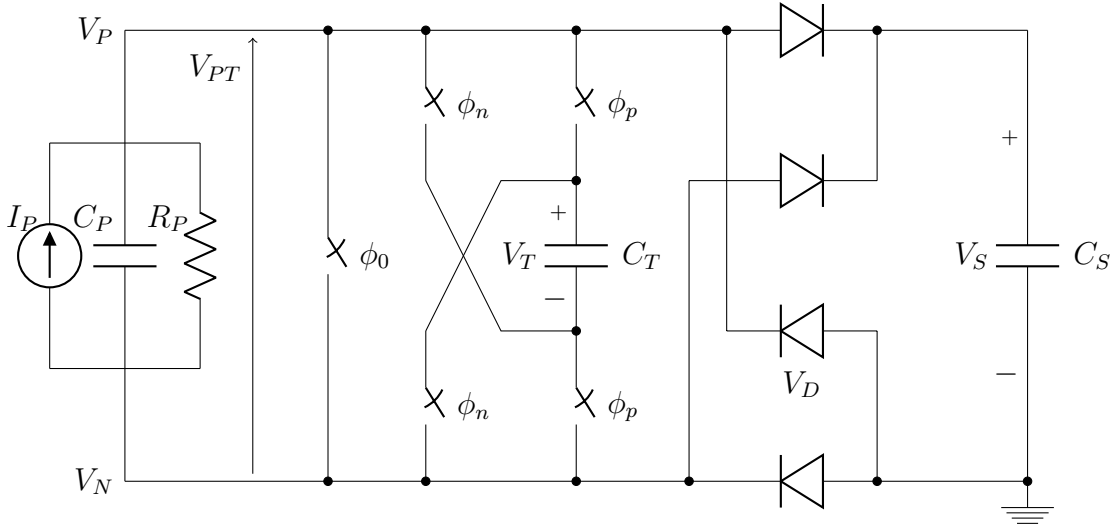
\begin{figure} []
\centering
\begin{tikzpicture}[scale = 1]

\draw (-1,7)  -- (-1,5.5) to[C]  (-1,2.5) --(-1,1) (-1,4.5) node[anchor=east]{$C_P$};
\draw (-2,2.5) to[I]  (-2,5.5) (-2,4.5) node[anchor=east]{$I_P$}; 
\draw (-0,5.5) to[R]  (-0,2.5) (-0,4.5) node[anchor=east]{$R_P$}; 
\draw (-2,5.5) -- (-0,5.5) (-2,2.5) -- (-0,2.5);

\draw (-1,7) node[anchor=east]{$V_P$}  (-1,1)node[anchor=east]{$V_N$};
\draw [->] (0.7,1.1) -- (0.7,6.9);  
\draw (0.7,6.5) node[anchor=east]{$V_{PT}$};

\draw (-1,7) -- (7,7) to [D] (9,7);
\draw (9,1) to [D] (7,1) -- (-1,1);
\draw (6.5,1) to [short,*-] (6.5,5) -- (7,5) to [D] (9,5) to [short,-*] (9,7);
\draw (9,1) to[short,*-] (9,3) to [D, l=$V_D$] (7,3) to [short,-*] (7,7);
\draw (9,7) -- (11,7) (9,1)--(11,1) ;
\draw (11,7) to [C=$C_S$, v=$V_S$] (11,1);

\draw (1.7,7) node[circ]{} to[cspst, l=$\phi_0$] (1.7,1) node[circ]{};
\draw (5,7) node[circ]{} to[cspst, l=$\phi_p$] (5,5) to[C=$C_T$, v=$V_T$] (5,3) to[cspst, l=$\phi_p$] (5,1) node[circ]{} ;
\draw (5,5) node[circ]{} -- (4,5) -- (3,3) to[cspst, l=$\phi_n$] (3,1) node[circ]{};
\draw (5,3) node[circ]{} -- (4,3) -- (3,5)  (3,7)node[circ]{} to[cspst, l=$\phi_n$] (3,5) ;
\draw (11,1) node[circ]{} node[ground]{};

\end{tikzpicture}
\caption{Proposed interface circuit}
\label{fig:1907_cdosiosc}
\end{figure}

Fig. \ref{fig:1907_cdosiosc} shows the design of a one-stage SSHC rectifier where a temporary charge storage switched capacitor $C_T$ is used to synchronously flip the voltage $V_{PT}$. In order to perform the charge inversion, five analogue switches driven by three pulse signals ($\phi_p$, $\phi_0$ and $\phi_n$) are used. The three non-overlapping switching signals are synchronously generated to turn ON the five switches sequentially in a given order. The order of the three pulses depends on the polarization of the voltage $V_{PT}$. 

\begin{figure}
\centering
\includegraphics[width=0.9\linewidth]{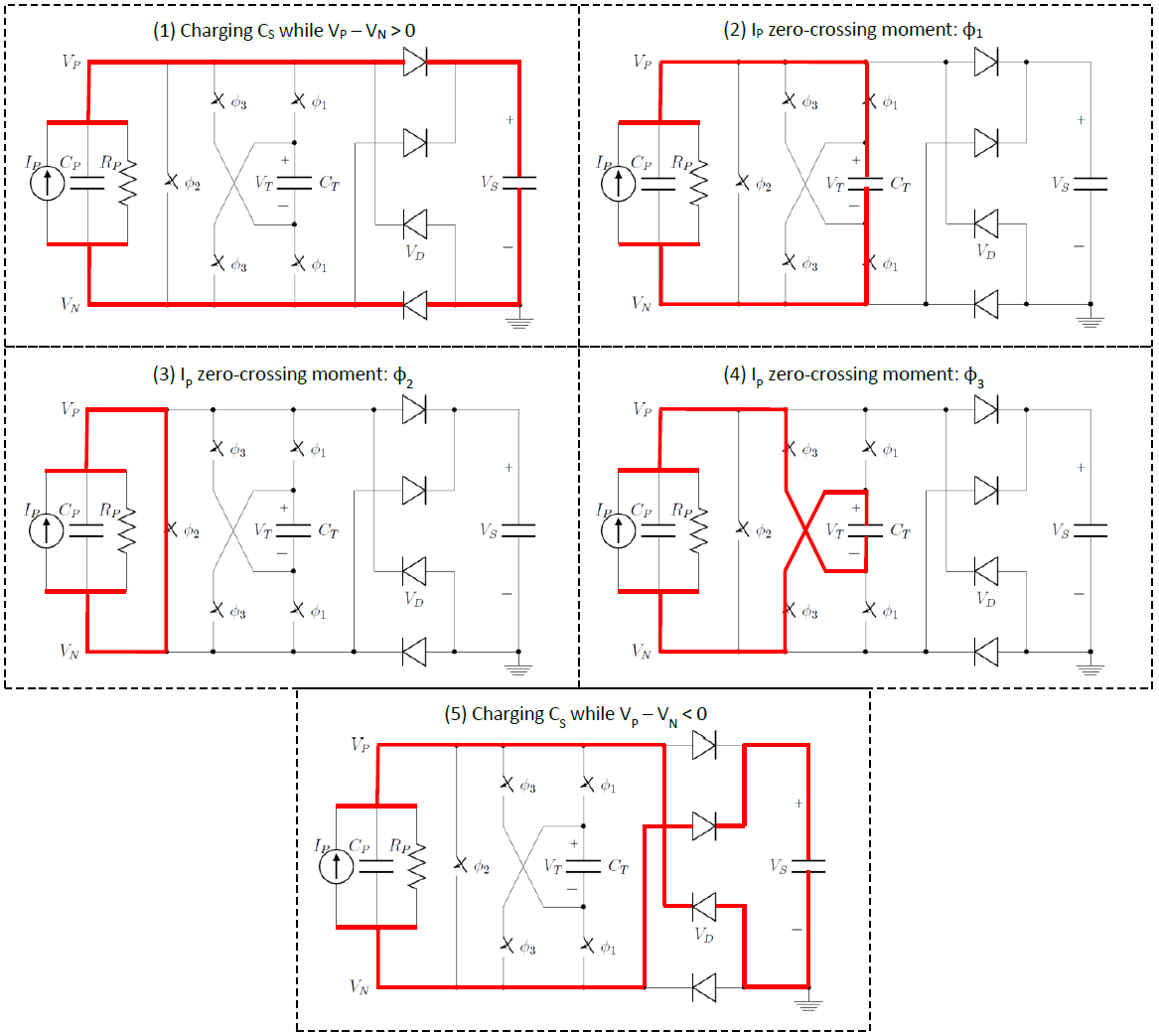}
\caption{Working principle of proposed SSHC rectifier to perform }
\label{fig:1907_wpopshcrtp}
\end{figure}

Fig. \ref{fig:1907_wpopshcrtp} shows the working principle of the one-stage SSHC rectifier. Before $I_P$ reaches zero-crossing point, the generated charge by the piezoelectric transducer flows into the storage capacitor $C_S$, as shown in first sub-figure. The polarization of voltage across the piezoelectric transducer is assumed that $V_{PT} > 0$; hence the top and the bottom diodes are conductive and $V_{PT} = V_S + 2V_D$ during this time. At the moment of $I_P$ zero-crossing point, a pulse $\phi_p$ is generated to let some charge in $C_P$ flow into $C_T$. At the next phase, $\phi_0$ turns ON the switch across the piezoelectric transducer and clears the remaining charge in $C_P$. At the phase of $\phi_n$, $C_T$ is connected to the piezoelectric transducer in an opposite sense, hence $V_{PT}$ goes to a negative value and a part of charge is inverted as a result. After the phase $\phi_n$, the polarization of $I_P$ changes and $V_{PT}$ goes to $-(V_S+2V_D)$ until the middle two diodes become conductive to start charging $C_S$ again. In the voltage inversion process shown in Fig. \ref{fig:1907_wpopshcrtp}, the order of the three signals is $\phi_p \rightarrow \phi_0 \rightarrow \phi_n$ because $V_P > V_N$ before the zero-crossing moment and $V_{PT}$ is aimed to be inverted from $V_S + 2V_D$ to a negative value. While in the other case of $V_P < V_N$, the order of the three signals should be $\phi_n \rightarrow \phi_0 \rightarrow \phi_p$. 

\begin{figure}
\centering
\begin{subfigure}{0.49\textwidth}
  \centering
  \includegraphics[width=1\linewidth]{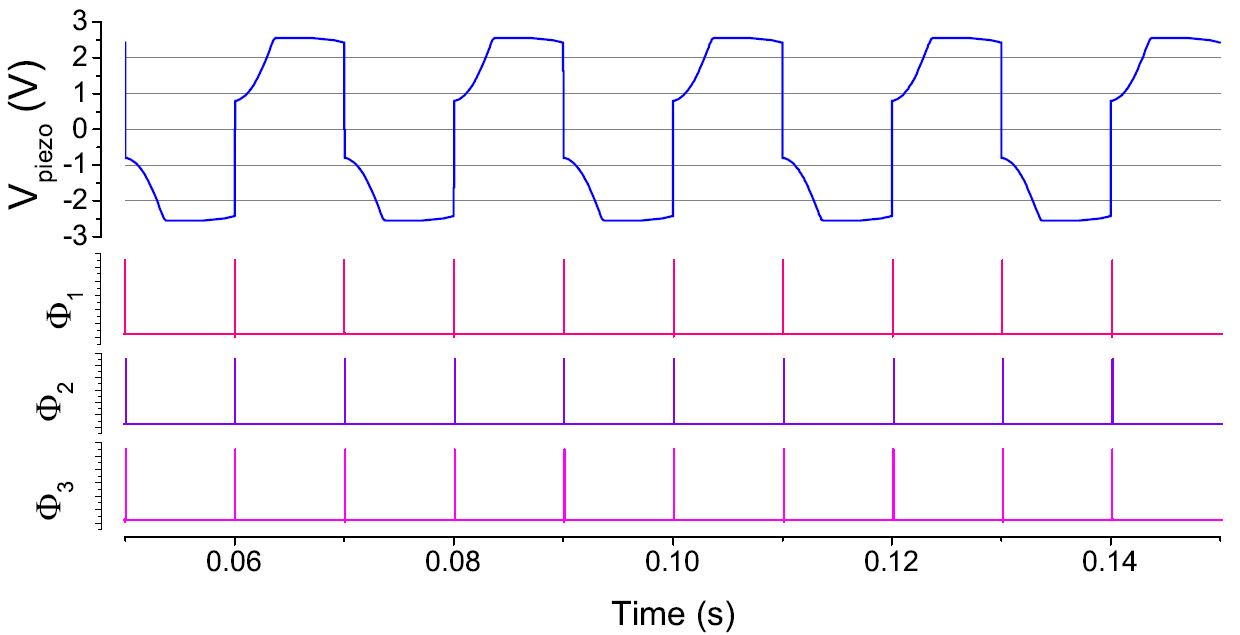}
  \caption{Simulated waveforms of $V_{PT}$, $\phi_p$, $\phi_0$ and $\phi_n$}
  \label{fig:1907_simwovpzphi}
\end{subfigure}
\begin{subfigure}{0.49\textwidth}
  \centering
  \includegraphics[width=1\linewidth]{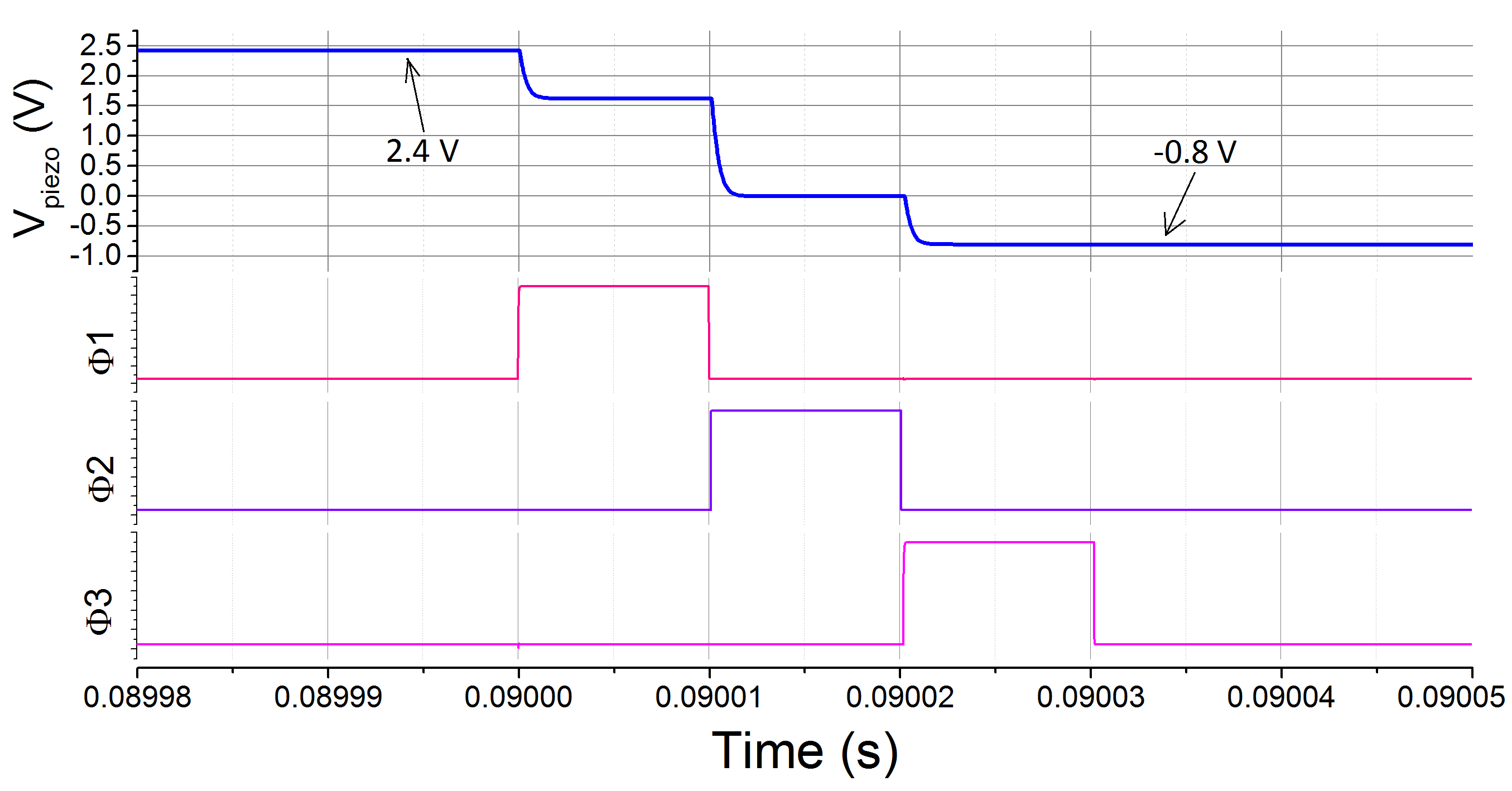}
  \caption{Zero-crossing moment while $V_{PT}$ is inverted from `+' to `-'}
  \label{fig:1907_simuvpzphialdwn}
\end{subfigure}
\begin{subfigure}{0.49\textwidth}
  \centering
  \includegraphics[width=1\linewidth]{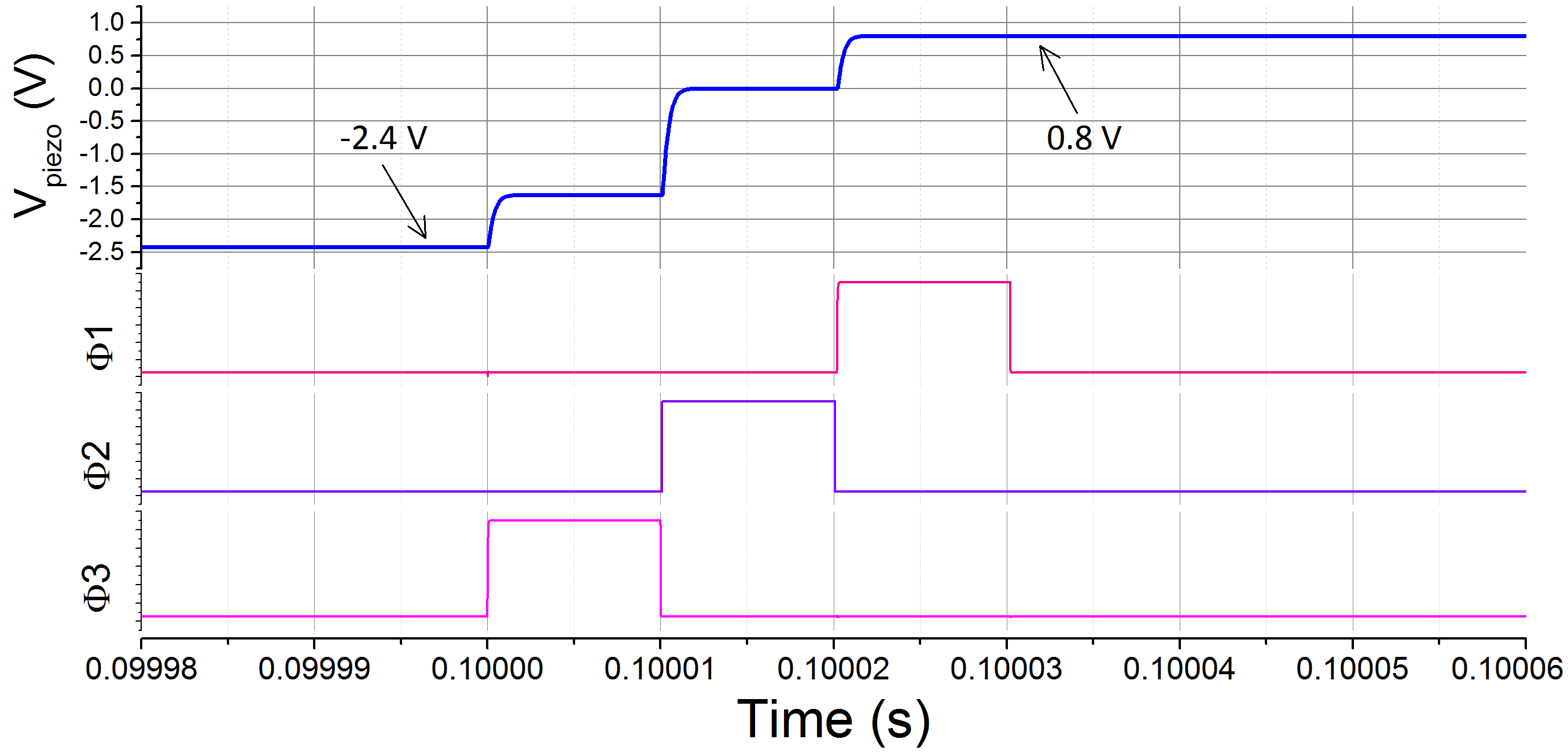}
  \caption{Zero-crossing moment while $V_{PT}$ is inverted from `-' to `+'}
  \label{fig:1907_simuvpzphialup}
\end{subfigure}
\caption{Simulated waveforms while $V_{PT}$ is being inverted}
\label{fig:1907_swwvpzibid}
\end{figure}

Fig. \ref{fig:1907_simwovpzphi} shows the simulated waveforms of $V_{PT}$, $\phi_p$, $\phi_0$ and $\phi_n$ for five periods of $I_P$. For each zero-crossing moment, the three pulse signals are generated sequentially and it can be seen that $V_{PT}$ is partially inverted. Fig. \ref{fig:1907_simuvpzphialdwn} presents the voltage inversion in more detail while $V_{PT}$ is inverted from positive to negative. In this case, the switch $\phi_p$ is first turned ON and the capacitors, $C_P$ and $C_T$ are connected in the same polarization. From the waveform of $V_{PT}$, it can be seen that it goes down a little because the charge in $C_P$ is shared between the $C_P$ and $C_T$. At the $\phi_0$ phase, $C_P$ is shorted and the remaining charge in it is cleared; hence $V_{PT}$ goes to \SI{0}{V}. At the $\phi_n$ phase, $C_T$ and $C_P$ are connected in an polarization opposite to the phase $\phi_p$. At this state, some charge in $C_T$ flows into $C_P$ until they have the same voltage values across them and $V_{PT}$ goes to a negative value as a result. In the simulation, $V_{PT}$ equals to \SI{2.4}{V} before the zero-crossing moment and it goes to \SI{-0.8}{V} after the inversion process. Fig. \ref{fig:1907_simuvpzphialup} shows waveforms while $V_{PT}$ is inverted from negative to positive and the order that the three pulse signals to be generated is $\phi_n \rightarrow \phi_0 \rightarrow \phi_p$. In addition to the order, the three signals should also be non-overlapping to avoid any unwanted charge flow.

\subsection{Performance analysis with $C_T = C_P$} \label{1907_pefays}

As discussed, the SSHC rectifier is able to invert some charge stored in $C_P$ and it is useful to calculate how much charge is inverted and the condition to achieve this performance. Before a zero-crossing moment, it is assumed that $V_{PT}$ is positive and it equals to $V_S + 2V_D$, noted as $V_0= V_S + 2V_D$ for simplicity. $C_T$ is assumed to have no charge at the beginning hence $V_T = \SI{0}{V}$. At the first zero-crossing moment of $I_P$, $\phi_p$ is first be turned ON because $V_{PT}$ is positive. $C_P$ and $C_T$ are connected and the charge flows into $C_T$ until the voltages across the two capacitors are equal. As the total charge keeps unchanged, the voltage across $C_P$ and $C_T$ at the end of the first phase is:

\begin{equation}\label{eq:1907_vpvtphi1}
\centering
V_{PT1} = V_{T1} = \frac{C_P}{C_P + C_T} V_{0}
\end{equation}

At the second phase, a pulse $\phi_0$ is generated. The remaining charge in $C_P$ is cleared and the charge in $C_T$ is unchanged. Hence, the voltage across $C_P$ and $C_T$ at the end of the second phase is:

\begin{equation}\label{eq:1907_vpvtphi2}
\centering
\begin{split}
V_{PT2} &= 0 \\
V_{T2} &= V_{T1} = \frac{C_P}{C_P + C_T} V_{0}
\end{split}
\end{equation}

At the phase $\phi_n$, $C_T$ is connected with $C_P$ again in an opposite sense to charge $C_P$ to a negative voltage. As the total charge in the two capacitors is the remaining charge in $C_T$ shown in (\ref{eq:1907_vpvtphi2}), hence the voltages of $V_{PT}$ and $V_T$ at the end of this phase are: 

\begin{equation}\label{eq:1907_vpvtphi3}
\centering
\begin{split}
V_{PT3} = - V_{T3} = - \frac{C_P C_T}{(C_P + C_T)^2} V_{0}
\end{split}
\end{equation}

It can be seen that $V_{PT}$ is a negative value at the end of the zero-crossing moment. By setting the derivative of the expression in (\ref{eq:1907_vpvtphi3}) at 0, it can be found that $V_{PT3}$ attains its minimum value while $C_T = C_P$. Therefore, the minimum value of $V_{PT}$ at the end of the first charge inversion is:

\begin{equation}\label{eq:1907_bvpvtphi3}
\centering
V_{PT3} = - V_{T3} = -\frac{1}{4} V_{0} \ \ \ \  ( while\  C_P = C_T )
\end{equation}

The resulting voltage in (\ref{eq:1907_bvpvtphi3}) is obtained after the first charge inversion and the initial voltage across $C_T$ is assumed at \SI{0}{V} at the beginning. However, before the second zero-crossing moment, $V_T$ is no longer \SI{0}{V}, but $\frac{1}{4} V_{0}$. $V_{PT}$ now equals to $-V_0$ because it needs to inverted from negative to positive. Assuming $C_T = C_P$ is chosen for future calculations, $V_{PT}$ and $V_T$ values after each phase of $\phi_n$, $\phi_0$ and $\phi_p$ at the second charge inversion stage are:

\begin{equation}\label{eq:1907_vpzszp}
\centering
\begin{split}
before\ \phi_n: \ \ & V_{PT} = - V_{0}, \ \ V_T = \frac{1}{4} V_{0} \\
\Rightarrow after\ \phi_n: \ \ & V_{PT} = -V_T = - (\frac{1}{4} + 1) \frac{1}{2} V_0 \\
\Rightarrow after\ \phi_0: \ \ & V_{PT} = \SI{0}{V}, \ \ V_T = (\frac{1}{4} + 1) \frac{1}{2} V_0\\
\Rightarrow after\ \phi_p: \ \ & V_{PT} = V_T = (\frac{1}{4} + 1) \frac{1}{4} V_0 = ((\frac{1}{4})^2 + \frac{1}{4}) V_0 = \frac{5}{16} V_0
\end{split}
\end{equation}

As $\frac{5}{16} > \frac{1}{4}$, more charge is inverted in the second zero-crossing moment compared to the first one. Due to the accumulation of remaining charge in $C_T$, the resulting $|V_{PT}|$ at the end of the $n^{th}$ inversion stage is:

\begin{equation}\label{eq:1907_vpateotn}
\centering
\begin{split}
|V_{PT}| &= ((\frac{1}{4})^n + \dots + (\frac{1}{4})^2 + \frac{1}{4}) V_0 = \sum_{\substack{1 \leq i \leq n}} (\frac{1}{4})^i V_0 = \frac{\frac{1}{4} - (\frac{1}{4})^n}{1-\frac{1}{4}} V_0 \\
\Rightarrow & \lim_{n \to \infty} |V_{PT}| = \frac{1}{3} V_0
\end{split}
\end{equation}

\begin{figure}
\centering
\includegraphics[width=0.6\linewidth]{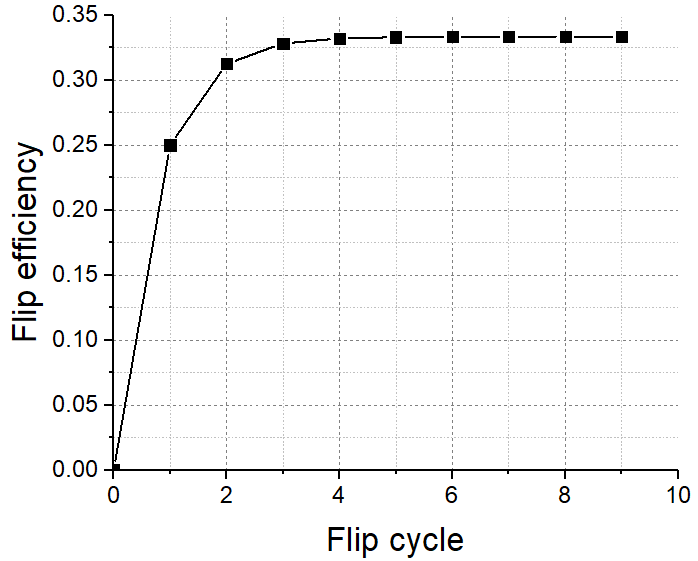}
\caption{Voltage flip efficiency for the first 10 flip cycles when $C_T = C_P$. }
\label{fig:1907_fliptec}
\end{figure}

While $n$ tends to infinity, $|V_{PT} |_{n \rightarrow \infty} = \frac{1}{3} V_0$, which means one third of charge can be inverted theoretically while $C_T= C_P$. Fig. \ref{fig:1907_fliptec} shows the simulated voltage flip efficiencies for the first 10 flip cycles when $C_T= C_P$. It can be seen that the flip efficiency goes fast to approach 1/3 after several cycles. 

\subsection{Performance analysis with $C_T >> C_P$} \label{1907_pefaysg}

Since $C_T$ is assumed to be off-chip implemented, it is possible to achieve  $C_T >> C_P$ with a tiny SMD capacitor for $C_T$ because $C_P$ is usually as small as several nF. When $C_T >> C_P$, the charge stored in $C_T$ will accumulate much slower. However, after sufficient number of flip cycles, voltage across $C_T$ will approach half of the voltage across $C_P$ before flipping starts. Since  $C_T$ is very large, the voltage across $C_T$ barely decreases after charge is dumped back from $C_T$ to $C_P$ to finish voltage flipping. In this case, the remaining voltage across $C_P$ keeps almost constant. Hence, the voltage flip efficiency can be as high as 50\% in this case. 

\begin{figure}
\centering
\includegraphics[width=0.6\linewidth]{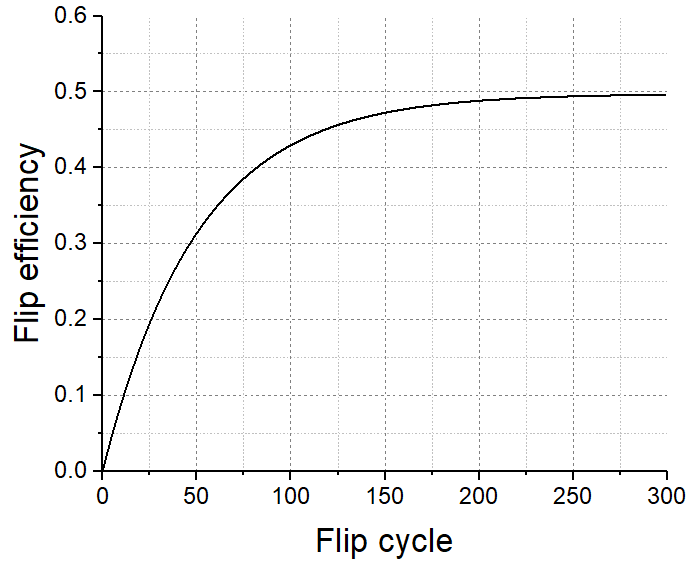}
\caption{Voltage flip efficiency for the first 300 flip cycles when $C_T = 100 C_P$. }
\label{fig:1907_fliptgc}
\end{figure}

Fig. \ref{fig:1907_fliptgc} shows the simulated voltage flip efficiencies for the first 300 flip cycles when $C_T >> C_P$. In this simulation, $C_T = 100 C_P$ is used. From this figure, it can be seen that the flip efficiency takes much more flip cycles to achieve the optimal level compared to the Fig. \ref{fig:1907_fliptec}. This is because $C_T$ is much larger now and it needs more charge from $C_P$ to charge $C_T$ to the sufficient level of voltage. The figure shows the voltage flip efficiency approaches 50\%, which is higher than 1/3, simulated for the case $C_T = C_P$.

\section{Conclusion} \label{1907_conclu}

An SSHC  (synchronized switch harvesting on capacitor) rectifier with one switched capacitor for piezoelectric vibration energy harvesters is analyzed in this paper. Similar to most of other highly-efficient active rectifiers (such as SSHI), it performs synchronous voltage inversion across the piezoelectric transducer every half cycle to minimize energy loss due to discharging and charging the internal capacitor. However, different from those rectifiers, the SSHC rectifier does not employ any inductor to perform the voltage inversion; it utilizes a capacitor instead. In previous studies, The capacitance of the employed capacitor is suggested to be equal to the internal capacitor of the piezoelectric transducer in order to achieve the performance to invert 1/3 of charge. In this paper, the voltage flip efficiency is simulated to be up to 50\% when the capacitance of the switched capacitor $C_T$ is much higher than the internal capacitor of the piezoelectric transducer $C_P$.


\bibliographystyle{IEEEtran}
\bibliography{./references}

\begin{thebibliography}{10}
\providecommand{\url}[1]{#1}
\csname url@samestyle\endcsname
\providecommand{\newblock}{\relax}
\providecommand{\bibinfo}[2]{#2}
\providecommand{\BIBentrySTDinterwordspacing}{\spaceskip=0pt\relax}
\providecommand{\BIBentryALTinterwordstretchfactor}{4}
\providecommand{\BIBentryALTinterwordspacing}{\spaceskip=\fontdimen2\font plus
\BIBentryALTinterwordstretchfactor\fontdimen3\font minus
  \fontdimen4\font\relax}
\providecommand{\BIBforeignlanguage}[2]{{%
\expandafter\ifx\csname l@#1\endcsname\relax
\typeout{** WARNING: IEEEtran.bst: No hyphenation pattern has been}%
\typeout{** loaded for the language `#1'. Using the pattern for}%
\typeout{** the default language instead.}%
\else
\language=\csname l@#1\endcsname
\fi
#2}}
\providecommand{\BIBdecl}{\relax}
\BIBdecl

\bibitem{Vullers2009}
R.~J.~M. Vullers \emph{et~al.}, ``Micropower energy harvesting,''
  \emph{Solid-State Electronics}, vol.~53, no.~7, pp. 684--693, 2009.

\bibitem{Rezaei2016atdom}
N.~Rezaei-Hosseinabadi \emph{et~al.}, ``A topology and design optimization
  method for wideband piezoelectric wind energy harvesters,'' \emph{IEEE
  Transactions on Industrial Electronics}, vol.~63, no.~4, pp. 2165--2173,
  2016.

\bibitem{Belleville2010}
M.~Belleville \emph{et~al.}, ``Energy autonomous sensor systems: Towards a
  ubiquitous sensor technology,'' \emph{Microelectronics Journal}, vol.~41,
  no.~11, pp. 740--745, 2010.

\bibitem{Niu2015auscsd}
S.~Niu \emph{et~al.}, ``A universal self-charging system driven by random
  biomechanical energy for sustainable operation of mobile electronics,''
  \emph{Nat Commun}, vol.~6, 2015.

\bibitem{Liangjunrui2012}
J.~Liang and W.-H. Liao, ``Improved design and analysis of self-powered
  synchronized switch interface circuit for piezoelectric energy harvesting
  systems,'' \emph{IEEE Transactions on Industrial Electronics}, vol.~59,
  no.~4, pp. 1950--1960, 2012.

\bibitem{luo2016daaombf}
Y.~Luo \emph{et~al.}, ``Design and analysis of a mems-based bifurcate-shape
  piezoelectric energy harvester,'' \emph{Aip Advances}, vol.~6, no.~4, p.
  045319, 2016.

\bibitem{Szarka2012}
G.~D. Szarka, B.~H. Stark, and S.~G. Burrow, ``Review of power conditioning for
  kinetic energy harvesting systems,'' \emph{Power Electronics, IEEE
  Transactions on}, vol.~27, no.~2, pp. 803--815, 2012.

\bibitem{Liu2012}
H.~Liu, C.~Lee, T.~Kobayashi, C.~J. Tay, and C.~Quan, ``Piezoelectric
  mems-based wideband energy harvesting systems using a frequency-up-conversion
  cantilever stopper,'' \emph{Sensors and Actuators A: Physical}, vol. 186,
  no.~0, pp. 242--248, 2012.

\bibitem{Blystad2010pmehs}
L.~C.~J. Blystad, E.~Halvorsen, and S.~Husa, ``Piezoelectric mems energy
  harvesting systems driven by harmonic and random vibrations,''
  \emph{Ultrasonics, Ferroelectrics, and Frequency Control, IEEE Transactions
  on}, vol.~57, no.~4, pp. 908--919, 2010.

\bibitem{Szarka2013upfabrfe}
G.~D. Szarka, S.~G. Burrow, and B.~H. Stark, ``Ultralow power, fully autonomous
  boost rectifier for electromagnetic energy harvesters,'' \emph{IEEE
  Transactions on Power Electronics}, vol.~28, no.~7, pp. 3353--3362, 2013.

\bibitem{Dongwon2014jasi}
D.~Kwon and G.~A. Rincon-Mora, ``A single-inductor 0.35 um cmos
  energy-investing piezoelectric harvester,'' \emph{IEEE Journal of Solid-State
  Circuits}, vol.~49, no.~10, pp. 2277--2291, 2014.

\bibitem{Sanchez2016apsrfp}
D.~A. Sanchez, J.~Leicht, F.~Hagedorn, E.~Jodka, E.~Fazel, and Y.~Manoli, ``A
  parallel-sshi rectifier for piezoelectric energy harvesting of periodic and
  shock excitations,'' \emph{IEEE Journal of Solid-State Circuits}, vol.~51,
  no.~12, pp. 2867--2879, 2016.

\bibitem{Ramadass2010}
Y.~K. Ramadass and A.~P. Chandrakasan, ``An efficient piezoelectric energy
  harvesting interface circuit using a bias-flip rectifier and shared
  inductor,'' \emph{IEEE Journal of Solid-State Circuits}, vol.~45, no.~1, pp.
  189--204, 2010.

\bibitem{Lefeuvre2017aopeh}
E.~Lefeuvre, A.~Badel, A.~Brenes, S.~Seok, M.~Woytasik, and C.~S. Yoo,
  ``Analysis of piezoelectric energy harvesting system with tunable sece
  interface,'' \emph{Smart Materials and Structures}, vol.~26, no.~3, p.
  035065, 2017.

\bibitem{Aktakka2014}
E.~E. Aktakka and K.~Najafi, ``A micro inertial energy harvesting platform with
  self-supplied power management circuit for autonomous wireless sensor
  nodes,'' \emph{IEEE Journal of Solid-State Circuits}, vol.~49, no.~9, pp.
  2017--2029, 2014.

\bibitem{badel2005efficiency}
A.~Badel, D.~Guyomar, E.~Lefeuvre, and C.~Richard, ``Efficiency enhancement of
  a piezoelectric energy harvesting device in pulsed operation by synchronous
  charge inversion,'' \emph{Journal of Intelligent Material Systems and
  Structures}, vol.~16, no.~10, pp. 889--901, 2005.

\end{thebibliography}

\end{document}